\documentclass[11pt]{article}
\usepackage{amssymb,amsmath,amsfonts}
\usepackage{graphicx}
\usepackage{graphics}
\usepackage{eepic,epsfig}
\usepackage{dcolumn}

\begin{document}

\title{On features of the radiation from an electron moving along a helix
inside a cylindrical hole in a homogeneous dielectric}
\author{S. R. Arzumanyan, L. Sh. Grigoryan, H. F. Khachatryan, \and A. S.
Kotanjyan, A. A. Saharian\thanks{%
Email address: saharian@ictp.it} \\
\textit{Institute of Applied Problems in Physics, National Academy of
Sciences} \\
\textit{25 Hr. Nersessian Street, 0014 Yerevan, Armenia}}
\date{\today}
\maketitle

\begin{abstract}
The radiation from a charge moving along a helical trajectory inside a
cylindrical hole in homogeneous dielectric medium is investigated. Prompted
by availability of materials with large dielectric permittivity $\varepsilon
$ and small absorption, we discuss the features of this type of radiation
for media with $\varepsilon \gg 1$. It is shown that there are high peaks in
the angular distribution of radiation intensity at well-defined harmonics.
The conditions are specified for the cavity-to-helix radii ratio, $\rho
_{1}/\rho _{0}$, under which the angle-integrated radiation intensity on
some harmonics exceeds that in the empty space. Though the amplification of
radiation intensity increases with increasing $\varepsilon $, the
corresponding "resonant" values of $\rho _{1}/\rho _{0}$ ratio are
practically independent of the dielectric permittivity of surrounding
medium. It is shown that an analogous amplification of radiation takes place
essentially for the same values of $\rho _{1}/\rho _{0}$ also for the
radiation in a cylindrical waveguide with conducting walls. An explanation
of this phenomenon is given.
\end{abstract}

\bigskip

PACS number(s): 41.60.Ap, 41.60.Bq


\bigskip

\section{Introduction}

\label{sec:int}

The radiation from a charged particle moving along a helical orbit in vacuum
has been widely discussed in literature (see, for instance, \cite%
{Bord99,Hofm04} and references given therein). This type of electron motion
is used in helical undulators for generating electromagnetic radiation in a
narrow spectral interval at frequencies ranging from radio or millimeter
waves to X-rays. The unique characteristics, such as high intensity and high
collimation, have resulted in extensive applications of this radiation in a
wide variety of experiments and in many disciplines. These applications
motivate the importance of investigations for various mechanisms of
controlling the radiation parameters. From this point of view, it is of
interest to consider the influence of a medium on the spectral and angular
distributions of the radiation.

In Ref. \cite{Saha05} we have investigated the radiation by a charged
particle moving along a helical orbit inside a dielectric cylinder immersed
into a homogeneous medium. Specifically, formulae are derived for the
electromagnetic fields and for the spectral-angular distribution of the
radiation intensity in the exterior medium. It is shown that under the
Cherenkov condition for dielectric permittivity of the cylinder and the
velocity of the particle image on the cylinder surface, strong narrow peaks
are present in the angular distribution for the number of radiated quanta.
At these peaks the radiated energy exceeds the corresponding quantity for a
homogeneous medium by some orders of magnitude. A special case of the
relativistic motion along the direction of the cylinder axis with
non-relativistic transverse velocity is discussed in detail and various
regimes for the undulator parameter are considered. The electromagnetic
fields generated inside the cylinder by the charge moving along a helical
orbit are considered in \cite{Saha06inside}. Note that the properties of the
radiation from a charged particle moving along a helical orbit in
homogeneous dispersive medium are studied in \cite{Gevo84}. In the present
paper we give the results of the further investigation of the radiation
features for a charge moving along a helix inside a cylindrical hole in a
homogeneous medium. We also compare this results with the radiation
properties in the case when the particle moves with helical trajectory
inside a cylindrical waveguide with conducting walls \cite{Kara77,Kota07}.

\section{Radiation from a charge moving inside a cylindrical hole in a
homogeneous medium}

\label{sec:Diel}

Consider a point charge $q$ moving along the helical trajectory of radius $%
\rho _{0}$ inside a cylindrical hole with radius $\rho _{1}$ in a
homogeneous medium with dielectric permittivity $\varepsilon $ (for
simplicity the magnetic permeability is taken to be unit). The particle
velocities along the axis of the hole (drift velocity) and in the
perpendicular plane we will denote by $v_{\parallel }$ and $v_{\perp }$,
respectively. In a properly chosen cylindrical coordinate system ($\rho
,\phi ,z$) the corresponding motion is described by the coordinates $\rho
=\rho _{0}$, $\phi =\omega _{0}t$, $z=v_{\parallel }t$, where the $z$-axis
coincides with the hole axis and $\omega _{0}=v_{\perp }/\rho _{0}$ is the
angular velocity of the charge. This type of motion can be produced by a
uniform constant magnetic field directed along the axis of a cylinder, by a
circularly polarized plane wave, or by a spatially periodic transverse
magnetic field of constant absolute value and a direction that rotates as a
function of the coordinate $z$.

In \cite{Saha05} it have been investigated the radiation by a charged
particle moving along a helical orbit inside a dielectric cylinder immersed
into a homogeneous medium. Now we turn to the investigation of the features
of the radiation intensity to the exterior medium at a given harmonic $m$ in
the case when the particle moves along helical trajectory inside a
cylindrical hole in dielectric medium with the dielectric permittivity $%
\varepsilon \gg 1$. At large distances from the charge trajectory the
dependence of elementary waves on the space-time coordinates has the form $%
\exp [i\omega _{m}\sqrt{\varepsilon }(\rho \sin \vartheta +z\cos \vartheta
-ct/\sqrt{\varepsilon })/c]$ which describes the wave with the frequency

\begin{equation}
\omega _{m}=\frac{m\omega _{0}}{|1-v_{\parallel }\sqrt{\varepsilon }\cos
\vartheta /c|},  \label{omegamtet}
\end{equation}%
propagating at the angle $\vartheta $ to the $z$-axis. The number of the
radiated quanta per period of a particle rotation for the harmonic $m\neq 0$
is given by the formula (see \cite{Saha05})
\begin{equation}
\frac{dN_{m}}{d\Omega }=\frac{q^{2}m\sqrt{\varepsilon }}{\pi ^{2}\hbar
c|1-v_{\parallel }\sqrt{\varepsilon }\cos \vartheta /c|^{2}}\left\{
\left\vert D_{m}^{(1)}-D_{m}^{(-1)}\right\vert ^{2}+\left\vert
D_{m}^{(1)}+D_{m}^{(-1)}\right\vert ^{2}\cos ^{2}\vartheta \right\} .
\label{Int3}
\end{equation}%
In (\ref{Int3}), $d\Omega =\sin \vartheta d\vartheta d\phi $ is the solid
angle element and

\begin{eqnarray}
D_{m}^{(p)} &=&\frac{v_{\perp }}{c\rho _{1}W_{m+p}}\left[ J_{m+p}(\lambda
_{0}\rho _{0})+\frac{p\lambda _{1}}{2\alpha _{m}}J_{m+p}(\lambda _{0}\rho
_{1})H_{m}(\lambda _{1}\rho _{1})\sum_{l=\pm 1}\frac{J_{m+l}(\lambda
_{0}\rho _{0})}{W_{m+l}}\right]  \notag \\
&&+\frac{v_{\parallel }J_{m}(\lambda _{0}\rho _{0})}{c\rho _{1}W_{m}}\left[ p%
\frac{\lambda _{1}J_{m+p}(\lambda _{0}\rho _{1})H_{m}(\lambda _{1}\rho _{1})%
}{\alpha _{m}W_{m+p}\tan \vartheta }-\tan \vartheta \right] .  \label{Dmp}
\end{eqnarray}%
In these formulae $J_{m}(x)$, $H_{m}(x)\equiv H_{m}^{(1)}(x)$ are the Bessel
and Hankel functions, the coefficients $\lambda _{j}$ are defined by the
expressions
\begin{equation}
\lambda _{0}=\frac{m\omega _{0}}{c}\frac{\sqrt{1-\varepsilon \cos
^{2}\vartheta }}{1-v_{\parallel }\sqrt{\varepsilon }\cos \vartheta /c},\text{
}\lambda _{1}=\frac{m\omega _{0}}{c}\frac{\sqrt{\varepsilon }\sin \vartheta
}{1-v_{\parallel }\sqrt{\varepsilon }\cos \vartheta /c},  \label{lambondb}
\end{equation}%
and
\begin{eqnarray}
W_{m} &=&J_{m}(\lambda _{0}\rho _{1})\frac{\partial H_{m}(\lambda _{1}\rho
_{1})}{\partial \rho _{1}}-H_{m}(\lambda _{1}\rho _{1})\frac{\partial
J_{m}(\lambda _{0}\rho _{1})}{\partial \rho _{1}},  \label{Wm} \\
\alpha _{m} &=&\frac{1}{\varepsilon _{1}-1}-\frac{\lambda _{0}J_{m}(\lambda
_{0}\rho _{1})}{2}\sum_{l=\pm 1}l\frac{H_{m+l}(\lambda _{1}\rho _{1})}{%
W_{m+l}}.  \label{alfam}
\end{eqnarray}

\begin{figure}[tbph]
\begin{center}
\epsfig{figure=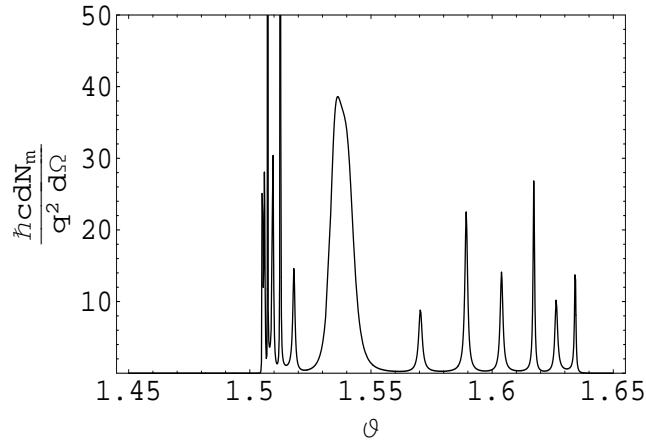,width=8.5 cm,height=6. cm}
\end{center}
\caption{The angular density of the number of the radiated quanta, $(\hbar
c/q^{2})dN_{m}/d\Omega $, as a function of the angle $\protect\vartheta $
for the harmonic $m=1,$ $E=10$ $MeV,$ $\protect\varepsilon =230,$ $\protect%
\phi _{0}=\protect\pi /3,$ $\protect\rho _{1}/\protect\rho _{0}=11.726.$}
\label{fig1}
\end{figure}

In figure \ref{fig1} we have presented the dependence of the angular density
of the number of the radiated quanta per period of an electron rotation, as
a function of the angle $\vartheta $ for the harmonic $m=1$. The values for
the other parameters are as follows: electron energy $E=10$ MeV, dielectric
permittivity of the surrounding medium $\varepsilon =230$, and $\phi
_{0}=\pi /3$, $\rho _{1}/\rho _{0}=11.726$. Here we have introduced the
angle $\phi _{0\text{ }}$between the particle velocity and the axis $z$:
\begin{equation}
v_{\perp }=v\sin \phi _{0},\text{ }v_{\parallel }=v\cos \phi _{0}.
\label{veloc}
\end{equation}

\begin{figure}[tbph]
\begin{center}
\epsfig{figure=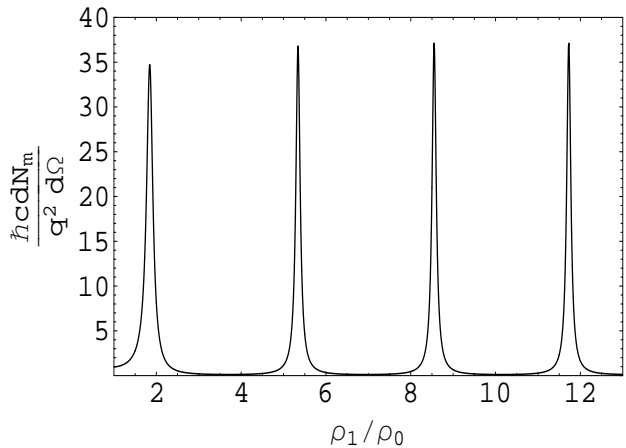,width=8.5 cm,height=6. cm}
\end{center}
\caption{The angular density of the number of the radiated quanta, $(\hbar
c/q^{2})dN_{m}/d\Omega $, as a function of the ratio $\protect\rho _{1}/%
\protect\rho _{0}$ for $\protect\vartheta _{0}=1.53786$. The values of the
other parameters are the same as those for figure \protect\ref{fig1}.}
\label{fig2}
\end{figure}

In the discussion below (see equation (\ref{coherent})) we will introduce
also the angle $\phi _{1}$ for which the velocity of the propagation for the
electromagnetic waves along the cylinder axis coincides with the projection
of the particle velocity on this axis. From the graph it is seen that,
though there are strong narrow peaks in left part of the graph (on the
location of these peaks see \cite{Saha05}), the main contribution to the
number of radiated quanta integrated over the angle $\vartheta $, comes from
the peak near $\vartheta _{0}=1.53786$. In the reference frame moving along
the $z$-axis by the velocity $v_{\parallel }$, this angle corresponds to the
angle $\vartheta _{0}^{\prime }=\pi /2$. It is related to the angle $\phi
_{1}$ by the formula
\begin{equation}
\vartheta _{0}=\arccos (\cos \phi _{1}/\sqrt{\varepsilon }).  \label{tet0}
\end{equation}%
In figure \ref{fig2} we give the angular density of the number of the
radiated quanta on the harmonic $m=1$ as a function of the ratio $\rho
_{1}/\rho _{0}$ for $\vartheta _{0}=1.53786$. The values of the other
parameters are the same as those for figure \ref{fig1}.

\begin{figure}[tbph]
\begin{center}
\epsfig{figure=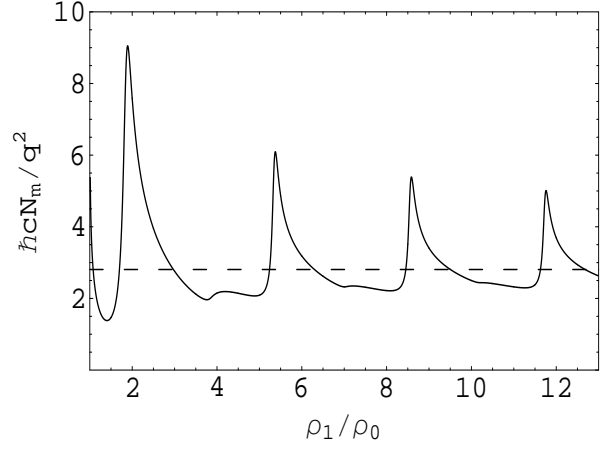,width=8.5 cm,height=6. cm}
\end{center}
\caption{The number of the radiated quanta, $(\hbar c/q^{2})N_{m}$, as a
function of the ratio $\protect\rho _{1}/\protect\rho _{0}$ for $m=1$ . The
values of the other parameters are as follows: $E=10$ $MeV$, $\protect\phi %
_{0}=\protect\pi /3$, $\protect\varepsilon =230$.}
\label{fig3}
\end{figure}

Integrating $dN_{m}$ over the angles, we obtain the total number of the
radiated quanta on a given harmonic. In figure \ref{fig3} we have presented
the number of radiated quanta $N_{m}$ for $m=1$ as a function of the ratio $%
\rho _{1}/\rho _{0}$. The values of the other parameters are as follows: $%
E=10$ MeV, $\phi _{0}=\pi /3$, $\varepsilon =230$. The horizontal straight
line corresponds to the number of the radiated quanta in the vacuum ($%
\varepsilon =1$). As it is seen from figure \ref{fig3} there are peaks in
the number of the radiated quanta for certain values of the ratio $\rho
_{1}/\rho _{0}$. The distance between the neighboring peaks, $\Delta (\rho
_{1}/\rho _{0})_{\mathrm{peak}}$, is approximately the same. The numerical
calculations have shown that similar features take place for other values of
the dielectric permittivity with the same locations of the peaks. The height
of the peaks is a decreasing function of $\varepsilon $. For example, the
height of the peak at $\rho _{1}/\rho _{0}\approx 1.89$ is $3.502$ for $%
\varepsilon =3$.

\begin{figure}[tbph]
\begin{center}
\epsfig{figure=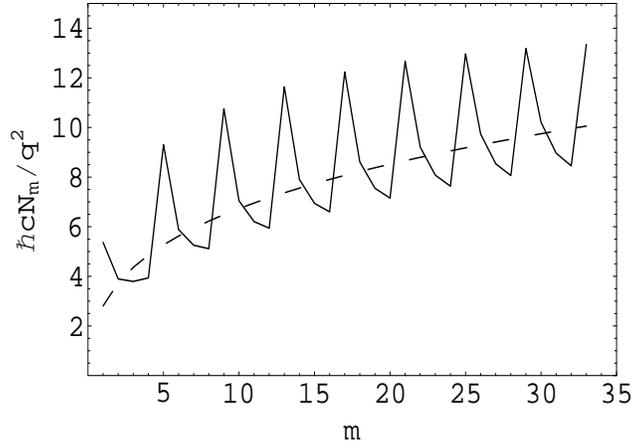,width=8.5 cm,height=6. cm}
\end{center}
\caption{The number of radiated quanta, $(\hbar c/q^{2})N_{m}$, as a
function of the number of harmonic $m$ for the values of the parameters $%
\protect\rho _{1}/\protect\rho _{0}=8.595$, $E=10$ MeV, $\protect\phi _{0}=%
\protect\pi /3$, $\protect\varepsilon =230$.}
\label{fig4}
\end{figure}

From the numerical calculations it follows that the amplification of the
radiation takes place at once on a set of harmonics. In figure \ref{fig4} we
have plotted the dependence of the number of radiated quanta on the number
of harmonic $m$ for the values of the parameters $\rho _{1}/\rho _{0}=8.595$%
, $E=10$ MeV, $\phi _{0}=\pi /3$, $\varepsilon =230$. On the same graph we
have also plotted the number of radiated quanta in the vacuum (dashed
curve). We see the following regularity: if there are peaks in the
dependence of the number of radiated quanta as a function of the ratio $\rho
_{1}/\rho _{0}$ for $m=1$, then the peaks are present also for the harmonics
$m=4l+1$, $l=1,2,\ldots $.

\section{Radiation in a cylindrical waveguide}

\label{sec:Waveguide}

In this section we consider the radiation intensity for a charge moving in a
helical orbit inside an empty circular waveguide with radius $\rho _{1}$. We
will assume that the waveguide walls are perfectly conducting. The radiation
field inside the waveguide is a superposition of TM and TE waves. The
corresponding frequencies are equal
\begin{equation}
\omega _{m,n}^{(\sigma ,\pm )}=\frac{m\omega _{0}}{1-v_{\parallel }^{2}/c^{2}%
}\left[ 1\pm \frac{v_{\parallel }}{c}\sqrt{1-b_{m,n}^{(\sigma )2}\left(
1-v_{\parallel }^{2}/c^{2}\right) }\right] ,\;b_{m,n}^{(\sigma )}\equiv
\frac{cj_{m,n}^{(\sigma )}}{m\omega _{0}\rho _{1}}.  \label{ommnsigw}
\end{equation}%
In these formulae $\sigma =0,1$ correspond to TM and TE waves respectively, $%
j_{m,n}^{(\sigma )}$, $n=1,2,\ldots $, is the $n$th positive zero of the
Bessel function $J_{m}(x)$ ($\sigma =0$) and its derivative $J_{m}^{\prime
}(x)$ ($\sigma =1$). We have carried out numerical calculations for the
number of the quanta on a given harmonic radiated per one period of the
particle orbiting:%
\begin{equation}
N_{m}=\sum_{\sigma =0,1}\sum_{n=1}^{n_{\max }^{(\sigma )}}N_{m,n}^{(\sigma
)},\;m=1,2,\ldots .  \label{Int2}
\end{equation}%
The upper limit of the summation over $n$ in this formula is defined by the
relation
\begin{equation}
j_{m,n_{\max }^{(\sigma )}}^{(\sigma )}<\frac{m\omega _{0}\rho _{1}}{c\sqrt{%
1-v_{\parallel }^{2}/c^{2}}}<j_{m,n_{\max }^{(\sigma )}+1}^{(\sigma )}.
\label{defnmax}
\end{equation}%
For separate terms in (\ref{Int2}) one has the formulae \cite{Kota07}%
\begin{equation}
N_{m,n}^{(0)}=\frac{8\pi q^{2}m}{\hbar c}\frac{J_{m}^{2}(j_{m,n}^{(0)}\rho
_{0}/\rho _{1})}{j_{m,n}^{(0)2}J_{m+1}^{2}(j_{m,n}^{(0)})}\sqrt{%
1-b_{m,n}^{(0)2}\left( 1-v_{\parallel }^{2}/c^{2}\right) },  \label{ITEn}
\end{equation}

\begin{equation}
N_{m,n}^{(1)}=\frac{8\pi q^{2}}{\hbar cm(\rho _{1}/\rho _{0})^{2}}\frac{%
J_{m}^{^{\prime }2}(j_{m,n}^{(1)}\rho _{0}/\rho _{1})}{\left(
j_{m,n}^{(1)2}-m^{2}\right) J_{m}^{2}(j_{m,n}^{(1)})}\frac{j_{m,n}^{(1)2}}{%
\sqrt{1-b_{m,n}^{(1)2}\left( 1-v_{\parallel }^{2}/c^{2}\right) }}.
\label{ITMn}
\end{equation}%
If the condition%
\begin{equation}
j_{m,n}^{(\sigma )}\sqrt{1-v_{\parallel }^{2}/c^{2}}=m(v_{\perp }\rho
_{1})/(c\rho _{0})  \label{peakcond}
\end{equation}%
takes place then the intensity for the TE waves defined by formulae (\ref%
{ITEn}) goes to infinity. However, under these conditions the absorption in
the walls of the waveguide becomes important. We can introduce the angular
variable $\vartheta _{m,n}^{(\sigma ,\pm )}$ which is related to the
frequencies by the formula%
\begin{equation}
\omega _{m,n}^{(\sigma ,\pm )}=\frac{m\omega _{0}}{1-v_{\parallel }\cos
\vartheta _{m,n}^{(\sigma ,\pm )}/c}.  \label{omntet}
\end{equation}
It's possible values are determined by the formula
\begin{equation}
\cos \vartheta _{m,n}^{(\sigma ,\pm )}=\frac{b_{m,n}^{(\sigma
)2}v_{\parallel }/c\pm \sqrt{1-b_{m,n}^{(\sigma )2}\left( 1-v_{\parallel
}^{2}/c^{2}\right) }}{1+b_{m,n}^{(\sigma )2}v_{\parallel }^{2}/c^{2}}.
\label{costet}
\end{equation}%
Note that the singularity in the radiation intensity (\ref{ITMn}) noted
above corresponds to the values of the angular variable determined by the
condition%
\begin{equation}
\vartheta _{m,n}^{(\sigma ,\pm )}=\vartheta _{\perp },\;\vartheta _{\perp
}\equiv \arccos (v_{\parallel }/c).  \label{tetdiv}
\end{equation}%
In the reference frame moving along the direction of the axis $z$ with the
velocity $v_{\parallel }$ the angle corresponding to $\vartheta _{\perp }$
is equal to $\pi /2$.

\begin{figure}[tbph]
\begin{center}
\epsfig{figure=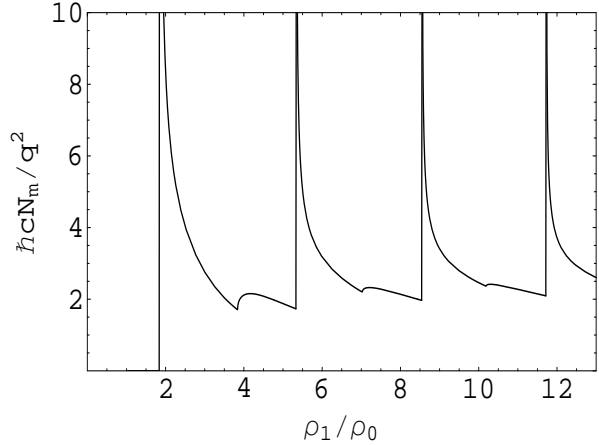,width=8.5 cm,height=6. cm}
\end{center}
\caption{The number of radiated quanta in the waveguide, $(\hbar
c/q^{2})N_{m}$, as a function of the ratio $\protect\rho _{1}/\protect\rho %
_{0}$. The values of the parameters are as follows: $m=1$, $E=10$ MeV, $%
\protect\phi _{0}=\protect\pi /3$.}
\label{fig5}
\end{figure}

In figure \ref{fig5} we have presented the number of quanta radiated on the
harmonic $m=1$ as a function of the ratio $\rho _{1}/\rho _{0}$ for the
values of the parameters: $E=10$ MeV, $\phi _{0}=\pi /3$.
Note that the location of the peaks in the number of radiated quanta is
determined by formula (\ref{peakcond}) with $\sigma =1$. Comparing the
numerical results presented in figures \ref{fig3} and \ref{fig5}, we see
that the location of the peaks in the dependence of the radiation intensity
of the ratio $\rho _{1}/\rho _{0}$ is the same for the problems considered
in sections \ref{sec:Diel} and \ref{sec:Waveguide}. An explanation of this
feature will be given in the next section.

\section{The reason of radiation amplification}

\label{sec:Expl}

\begin{figure}[tbph]
\begin{center}
\epsfig{figure=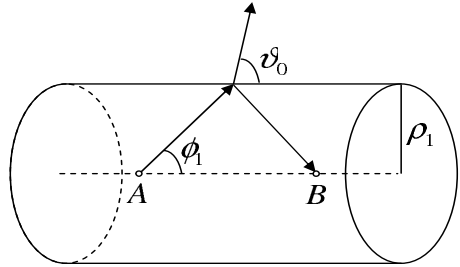,width=6. cm,height=3.5 cm}
\end{center}
\caption{On the explanation of the amplification for the radiation
intensity. }
\label{fig6}
\end{figure}

For simplicity let us consider the case when $\rho _{1}/\rho _{0}\gg 1$. In
this case for all harmonics the hole radius is much larger than the
wavelength of the radiation. We will assume that the particle is
relativistic. Obviously, there is a radiation direction $\phi _{1}$ for
which the electromagnetic wave radiated at the point A (see figure \ref{fig6}%
) after the reflection from the wall reaches the point B at the instant of
time when the particle is there. That condition is fulfilled if
\begin{equation}
v_{\parallel }=v\cos \phi _{0}=c\cos \phi _{1}.  \label{coherent}
\end{equation}%
For certain values of the phase this can lead to the essential change in the
radiation intensity. It is clear that the radiation corresponding to the
angle $\phi _{1}$ propagates in the exterior medium with the angle (\ref%
{tet0}). As it is seen from figure \ref{fig1}, exactly near $\vartheta _{0}$
intense radiation appears. The radiation intensity can be essentially
increased if the superposition of the waves at the point B is in the same
phase. This is possible if on the optical path $2\rho _{1}$ an integer
number of waves should be placed:%
\begin{equation}
\frac{2\rho _{1}}{\Lambda _{m}}+\frac{\Delta \varphi }{2\pi }=k,
\label{intcond}
\end{equation}%
where $\Delta \varphi $ is the phase change in the reflection, $\Lambda _{m}$
is the wavelength in the transverse direction, and $k=1,2,\ldots $. From (%
\ref{intcond}) and data given in figure \ref{fig2} one can conclude that the
change of the phase in the reflection of the wave%
\begin{equation}
\Delta \varphi \approx \pi /2\text{ at }\rho _{1}/\rho _{0}\gg 1.
\label{delta}
\end{equation}%
By taking into account the relation (\ref{coherent}) and formula (\ref%
{lambondb}), we have $\Lambda _{m}=2\pi c\sin \phi _{1}/(m\omega _{0})$, and
for the distance between two neighboring peaks one finds
\begin{equation}
\Delta (\rho _{1}/\rho _{0})\simeq \frac{\pi \tan \phi _{1}}{m\tan \phi _{0}}%
.  \label{ropeak}
\end{equation}%
So the peaks in figures \ref{fig2}, \ref{fig3} have to be equidistant, which
indeed takes place with relative accuracy $\approx 0.1\%$.

Now it is obvious from (\ref{intcond}), (\ref{delta}) that if a resonance
value for $\rho _{1}/\rho _{0}$ is found for some $m$, then for the same
values of the parameters the amplification of the radiation will be observed
for the harmonics $m^{\ast }=m(4l+1)$, $l=0,1,2,\ldots $, as well (see
figure \ref{fig4}).

As it is seen from formulae (\ref{intcond})-(\ref{ropeak}), for the problem
discussed in section \ref{sec:Diel} the resonance values for the ratio $\rho
_{1}/\rho _{0}$ do not depend on the dielectric permittivity of the
surrounding medium. For this reason, for the same values of $\rho _{1}/\rho
_{0}$ the increase of the radiation intensity has to be observed in the
waveguide as well. This is confirmed by the graphs given in figures \ref%
{fig3} and \ref{fig5}.

\section*{Acknowledgement}

The work has been supported by Grant No.~1361 from Ministry of Education and
Science of the Republic of Armenia.

\end{document}